\documentclass[12pt,preprint]{aastex}

\begin{document}
%



\title{A valid and fast spatial bootstrap for correlation functions}
\shorttitle{Spatial bootstrap for correlation functions}
\author{Ji Meng Loh}

\affil{Department of Statistics, Columbia University, New York, 10027,
USA}
\email{meng@stat.columbia.edu}

\begin{abstract}

In this paper, we examine the validity of non-parametric spatial
bootstrap as a procedure to 
quantify errors in estimates of $N$-point correlation functions.
We do this by means of a small simulation study with simple point
process models and estimating the two-point correlation
functions and their errors. The coverage of confidence intervals
obtained using bootstrap is compared with those obtained from assuming
Poisson errors. The bootstrap procedure considered here is adapted for
use with spatial (i.e.\ dependent) data. In particular, we describe a 
marked point bootstrap where, instead of resampling points or blocks
of points, we resample marks assigned to the data points. These marks
are numerical values that are based on the statistic of interest. 
We describe how the marks are defined for the two- and three-point
correlation functions. By
resampling marks, the bootstrap samples retain more of the dependence
structure present in the data. Furthermore, this method of bootstrap
can be performed much quicker than some other bootstrap methods for
spatial data, making it a more practical method with large datasets.
We find that with clustered point datasets, confidence intervals
obtained using the marked point bootstrap has empirical coverage
closer to the nominal level than the confidence intervals obtained
using Poisson errors. The bootstrap errors were also found to be
closer to the true errors for the clustered point datasets.

\end{abstract}


\section{Introduction} \label{sect:intro}

In analyses of survey data, such as those of galaxies or quasars, 
$N$-point correlation functions are often
estimated \citep[e.g.][]{kulkarni07, mccracken07, shen07}. These help to describe the structure of the observed
objects, such as the filamentary structure of galaxies that
has been observed, and to constrain the parameters of cosmological
models. Such estimates are only as important as their associated errors,
since it is the errors that indicate
the amount of agreement between two sets of data or between data and
simulations from a model.

In spatial point processes, the expressions for the standard errors of
correlation 
(and similar) functions have been worked out only for the simplest of
models. For example, \citet{ripley88} found approximations for the
variance of the $K$ function, an integral of the two-point correlation
function, for the Poisson process. These depend on such factors as the
shape of the observation region and the type of correction method for
boundary effects. \citet{landy93}, \citet{hamilton93} and others
worked out approximations for standard errors of various estimators of
the two-point correlation function under the Poisson or weakly
clustered models. Thus, very often, approximations such as Poisson
errors are used instead. However,
point data arising in astronomy are typically clustered and
non-Poisson. So while Poisson errors are useful and easy to compute,
they only serve as rough indications of the size of errors.

Besides using Poisson errors, errors can also be estimated by using
mock catalogs generated from a cosmological model. This method was
employed in \citet{eisenstein05}, where the initial conditions for the
cosmological model were selected independently. In the statistics
literature, this is referred to as parametric bootstrap, although the term
more commonly refers to mock datasets generated from the model using
parameter values fixed at the estimates from the data, instead of
being independently chosen.

An alternative is to use non-parametric bootstrap. This
involves generating new samples, called bootstrap samples, by
resampling from the actual data, and computing estimates for these new
samples. The distribution of these bootstrap estimates then serves as
a proxy for the actual distribution of the data estimates, so that
statistical inference, such as the construction of confidence
intervals, can be performed. Note that the procedure does not make any
specific model assumptions, thus the errors obtained by this method
can serve as a check of model assumptions.
Due to the simplicity and flexibility of the non-parametric bootstrap,
the method is attractive. What is desirable then is to make the
non-parametric bootstrap procedure work as well as possible for data
that is correlated, and check that it performs satisfactorily, so that
it can be useful as a tool in analysis. 

This paper thus examines the non-parametric bootstrap, specifically
bootstrap of spatial data, where the dependence present in the data is
of interest. There were some early misconceptions about how bootstrap
should be applied to spatial data. The naive method of resampling
individual points does not work in the spatial context.
In order for spatial bootstrap to be valid, the underlying
dependence structure has to be preserved as much as possible when
generating bootstrap samples.  Two common methods for doing this are the block
bootstrap and subsampling where blocks of data, instead of individual
data points, are resampled. We introduce these  methods
in Section \ref{sect:bootspatial} and describe their shortcomings.
In Section \ref{sect:improve} we describe the marked point bootstrap
\citep{loh02a} 
as a way to address these shortcomings. We describe how the marked
point bootstrap can be used with the two- and three-point correlation
function estimators, and by extension to estimators of $N$-point
correlation functions.
In Section \ref{sect:simstudy} we present results of a simulation
study using simple point process models comparing the empirical
coverage of confidence intervals obtained using non-parametric
bootstrap and using normal approximations with Poisson errors.

In this paper, we restrict ourselves to constructing nominal 95\%
confidence intervals, i.e. these confidence intervals are supposed to
contain the true value 95\% of the time. The empirical coverage of the
confidence intervals is the actual confidence level achieved by the
confidence intervals. In a simulation study with a known model, the
empirical coverage can be obtained by finding the number of confidence
intervals that contain the true value and then compared with the nominal
level. It is desirable, of course, for the empirical coverage to be
close to the nominal level. Furthermore, it is often better for the
empirical coverage to be higher instead of lower than the nominal
level, so that the procedure is conservative.

Bootstrap is a computationally intensive procedure. With the large
datasets now common in astronomy, even computing the $N$-point
correlation functions pose computational challenges. For example,
\citet{eisenstein05} avoided using the jackknife procedure for error
estimation because of the size of the data they used.
The way the
marked point bootstrap is formulated, however, makes it much faster than
subsampling (a generalization of the jackknife) and the
block bootstrap, so that applying the procedure to large datasets is
feasible as long as computing the actual estimates is
feasible. In Section \ref{sect:simstudy} we provide some
time measurements of the procedure used in our simulation study.

\section{Non-parametric bootstrap for spatial data}
\label{sect:bootspatial}

The non-parametric bootstrap was originally developed for independent data
\citep{efron94}. The main idea 
is to draw new samples from the actual data by sampling with
replacement a data point at a time. Bootstrap estimates of the same
statistic are computed from the bootstrap samples. With these
bootstrap estimates, confidence intervals, for example, can then be
constructed. This can be done in a variety of ways. Suppose $K$, $\hat{K}$ and
$\hat{K}^*_i, i=1, \ldots B$ are respectively the quantity of
interest, the estimate of $K$ computed from the data and the bootstrap
estimates, with $B$ equal to the number of bootstrap samples.
A simple method, called the basic
bootstrap interval in \citet{davison97}, is to set 
\begin{eqnarray}
& [2\hat{K}-\hat{K}^*_{(B+1)(1-\alpha/2)},
2\hat{K}-\hat{K}^*_{(B+1)\alpha/2} ] & \label{eqn:basicCI} 
\end{eqnarray}
as the $100(1-\alpha)$\% confidence interval for $K$. Here, $B$, the number
of bootstrap samples, is large, say, 999, and $\hat{K}^*_u$ is
the $u$-th ordered values of the bootstrap statistic. So, for example, with
$B=999$ bootstrap samples, a 95\% confidence interval for $K$ is given
by $[2\hat{K}-\hat{K}^*_{975},
2\hat{K}-\hat{K}^*_{25} ]$. In our simulation studies, we use
(\ref{eqn:basicCI}) to construct the confidence intervals. Standard
errors for $\hat{K}$ are estimated by the standard deviation of the
bootstrap estimates $\hat{K}^*$. 

While there are other methods of constructing confidence intervals
from bootstrap samples \citep[see][for example]{davison97}, the interest here is in the method of
generating the bootstrap samples, when the data is
spatial. \citet{snethlage99} rightly concludes that resampling
individual points do not work. If the resampled points are placed in
their original positions in the observation region, there will be
multiple points at single locations, which do not usually occur in most
data sets.  In their claim that bootstrap cannot be used for
analysis for clustering, \citet{simpson86} were also considering
bootstrap in terms of resampling individual points.

Due to the success of bootstrap for resampling independent data, it
has been extended to resample dependent data. Most of this work is for
time series, but can easily be applied to spatial data in two and
three dimensions. A common method is the block bootstrap: blocks of
the spatial data are sampled at random, then joined together to form a new
sample \citep{hall85, kunsch89, liu92}. Asymptotic arguments for the
validity of the bootstrap involve 
limiting the range of dependence, increasing the observation region size
and letting the resampling block size increase but at a slower rate
than the observation region size. In this asymptotic setting, we then
have many almost independent blocks of data, with each block itself
containing a large subsample \citep[see e.g.][]{lahiri03}.
However, the assumed conditions necessary to make the
calculations tractable also means that normal approximations work well too.
Some theoretical results show that the accuracy of bootstrap
estimates is of a higher order than the normal approximations. Whether
this difference is meaningful in actual practice is less clear. 
We believe that the role of non-parametric bootstrap is to serve as another
objective method to obtain standard errors that do not make any model
assumptions. Error estimates obtained using bootstrap can be used as a
way to assess or compare with other estimates of
errors.

\citet{kemball05} is a recent work in the astronomy literature
that examined bootstrap for dependent data. For the non-parametric
bootstrap, they focused on 
subsampling \citep{politis93a, politis99}, which can be considered 
a generalization of the jackknife procedure. In subsampling,
random portions of the data are deleted, and the remaining data are
treated as bootstrap samples. The standard deviation of the estimates computed
from these samples serves as an estimate of the error, less a factor
to adjust for the smaller subsamples and the large overlap. 

When estimating correlation functions, pairs or triplets etc of points
have to be counted. By joining independently resampled blocks together to
form the bootstrap sample, the block bootstrap creates artifical
configurations of points across the resampling blocks and distorts the
dependence structure in the data. This does not matter in asymptotic 
arguments because the effect becomes negligible if the range of the
correlation is fixed while the resampling blocks increase in
size. However, \citet{loh02a} found that the
actual coverage achieved by confidence intervals obtained using block
bootstrap can be much lower than the nominal percentage level for finite samples.

In subsampling, no artificial configurations of points are
created. However, while the correction weight accounts for the
difference in sample sizes between the bootstrap samples and the
actual data set, it does not account for the change in the boundary
effects due to the different resampling regions. Since subsampling
uses smaller regions as the bootstrap observation regions, boundary
effects are magnified. 
For subsampling, there is the temptation to use large
subsamples to try and retain more of the dependence structure, but
like block bootstrap, theoretical justification of the method requires
that the subsamples be small in size relative to the actual data set.
\citet{loh02a} also found that subsampling can
yield confidence intervals that attain very low empirical
coverage. They also found that the subsampling method is sensitive to
the fraction of the data used for subsampling.

\citet{loh02a}
proposed another version of spatial bootstrap, called marked point
bootstrap, that reduces the effect of joining independent blocks and
produces confidence intervals that achieve coverage closer to the
nominal level. This is described in the next section, where we also
show how it can be applied to the two- and three-point correlation
function estimators commonly used in astronomy.

\section{Improving the non-parametric bootstrap of spatial data}
\label{sect:improve}

Suppose $N$ points are observed in a region $A$. Furthermore, suppose
that the quantity of interest $K$ can be estimated using an estimator
of the form 
\begin{eqnarray}
\hat{K} & = & \frac{1}{N}\sum_{i=1}^N\sum_{j=1 \atop j\ne i}^N
\phi(x_i, x_j) \equiv \hat{\Phi}/N. 
\label{eqn:estimator}
\end{eqnarray}
Note that each point $i$ has an associated quantity $\sum_{j=1, j\ne i}^N
\phi(x_i,x_j)$, the inner sum of equation (\ref{eqn:estimator}).
Estimators of two-point statistics can be expressed in this form. In
this case, the
quantity $\phi(x_i,x_j)$ will depend on the distance
between $x_i$ and $x_j$. As an
aside, note that estimators of three-point statistics 
can be written in a similar form, with the inner sum replaced by a
double sum.

With point data, the term ``mark'' is used to refer to some additional
information associated with a point. This is usually some actual
measured value. For galaxy data, for example, marks could be
quantities such as luminosity, color and so on. In this paper, the
bootstrap method considered uses marks associated with the
points. However, these marks are not quantities such as luminosity
that are directly measured. Instead they are numerical quantities that
we construct and associate with the points. The actual values of these
marks are not random, but are constructed so that they relate to the
statistic that is of interest. If the statistic of interest is given
by equation (\ref{eqn:estimator}), then the mark associated with point
$i$, denoted by $m_i$, is equal to $\sum_{j=1, j\ne i}^N
\phi(x_i,x_j)$, so that\ $\hat{\Phi} = \sum_i m_i$. At the risk of
being repetitive, suppose that
$\hat{\Phi} \equiv DD(r) = \sum_{x\in D} \sum_{y\in D:y\ne x} 1\{ |x-y| \in
(r-dr,r+dr)\}$, for some $r$, is of
interest. This quantity is used in estimators of the
two-point correlation function, and is the number of pairs of points
separated by (roughly) distance $r$. Then the mark associated with
point $x$ is $\sum_{y\in D:y\ne x} 1\{ |x-y| \in (r-dr,r+dr)\}$, the
number of points that are roughly distance $r$ away from $x$. Note
that the sum of all the marks gives back the value of $DD(r)$. It is
also important to note that to compute the estimate
(\ref{eqn:estimator}), the marks have to be calculated anyway.
In regular applications, the algorithm doing the estimation does not 
individually record these marks, but keeps a running sum of the marks.
In order to do the marked point bootstrap, the
difference in terms of the code is that the marks now have to be
stored so that they can be used in the bootstrap step. 

In the block bootstrap, blocks of data
points are resampled and then joined together, forming a new dataset
from which $K$ is estimated using the new configuration of points that
was generated, yielding $\hat{K}^*$. In the marked point bootstrap,
blocks can be used to resample points as well. However, the crucial
difference is that the bootstrap estimate is computed, not from how
the resampled points are positioned, but from the marks that are associated
with these points. In other words, the marked point bootstrap
resamples the marks rather than the points and the bootstrap estimate is
computed by summing these resampled marks.

To be more precise, suppose that $N^*$ number of points have been resampled,
with the resampled points denoted by $x_j^*, j=1, \ldots
N^*$. Associated with each $x_j^*$ is a mark $m_{j^*}$. We denote this
mark  $m_{j^*}$ rather than  $m_j^*$ to emphasize the fact that these
marks are sampled from the actual data, i.e.\ computed from the
original dataset and not from the bootstrap sample. Then the bootstrap
estimate of $K$ is given by the average of the resampled marks: $$\hat{K}^*
= \hat{\Phi}^*/N^* = 
\sum_{j=1}^{N^*} m_{j^*}/N^*,$$ just like $\hat{K}$ is given by the average of
the actual marks. Note that in an actual implementation of the
procedure, all that is required is keeping track of how many times each
point is resampled.
The step-by-step procedure for estimating and resampling the
quantity (\ref{eqn:estimator}) is as follows:

\begin{enumerate}
\item For each point $i$, calculate $m_i = \sum_{j=1, j\ne i}^N
  \phi(x_i,x_j)$.
\item Obtain the estimate $\hat{K}$, using $\hat{K}= \sum_i m_i/N$.
\item Resample the points. This can be done by randomly placing blocks
  on to the observation region and keeping track of which point is
  resampled. Suppose point $i, i=1,\ldots , N$ is resampled $n^*_i$ times, and
  $N^*=\sum_i n_i^*$.
\item The bootstrap estimate is then $\hat{K}^* = \sum_i (n^*_i \times
  m_i)/N^*$
\item Repeat steps 3 and 4 to get $B$ bootstrap estimates.
\item Construct a confidence interval using (\ref{eqn:basicCI}).
\end{enumerate}

A few remarks about the procedure are in order.

\noindent {\bf Remark 1} Instead of randomly placing blocks, the observation region can
  be divided into a number of subregions, and the regions selected
  randomly with replacement. This latter method is sometimes referred
  to as using fixed blocks as opposed to moving blocks. It
  is generally considered that the moving blocks bootstrap works
  better in terms of convergence rates in asymptotic arguments.

\noindent {\bf Remark 2} The number of blocks used is so that the total area/volume of
  the blocks is equal to the original area/volume of the observation
  region. Note that in this case $N^*$ would usually not be equal to
  $N$, though they will be of the same order of magnitude. However,
  this does not pose problems since the statistic $\hat{K}$ is a mean
  of the marks.

\noindent {\bf Remark 3} There is no real consenus on the size of the resampling blocks
  to use. \citet{buhlmann99} did some work on determining the optimal block size
  from data. Intuitively, the procedure needs large blocks so that
  the correlation structure is less distorted, and a large enough
  number of blocks so that there is enough variability between
  bootstrap samples. If $K$ represents the number of blocks and $N$
  the number of data points (which is assumed to increase with the
  observation region size), theoretical work in e.g.\ \citet{lahiri03}
  suggests that consistency is
  achieved as $K\to \infty$ and $N/K \to \infty$. Thus some trade-off
  is needed. A rule-of-thumb is to divide each dimension
  of the observation region into at least three parts, i.e.\ nine
  blocks in 2D, 27 blocks in 3D. This would ensure enough variability
  between bootstrap samples. Of course, for correlation functions, the
  maximum value of the separation distance $r$ at which these
  functions are estimated would influence the
  decision on block size.
Fortunately, \citet{loh02a} found that the marked point bootstrap is less
  sensitive to block size than block bootstrap or subsampling: they
  resampled an absorber catalog using slices of the sphere and found that
  the bootstrap errors were similar for different sizes of the
  slices. Our simulation results also show little difference due to
  different block size.

There are a few advantages to this form of spatial bootstrap over the
regular block bootstrap. Since the bootstrap estimates are based on
the resampled marks and not on marks recalculated from the bootstrap
sample, the contribution to the bootstrap estimate is due to actual
pairs of points in the original dataset. This helps to minimize the
distortion of dependence structure in the dataset due to
resampling. 

Furthermore, for any block of resampled points,
information about the points just outside the block (and therefore not
sampled by this particular block) is captured by the marks associated
with the points that are sampled by the block. This helps to reduce
the variability of bootstrap results due to the size of the
resampling blocks, compared to block bootstrap or subsampling.
Also, since the resampling blocks do not need to be
joined together to form a contiguous region for the bootstrap sample,
there is flexibility in the choice of the shape of the resampling regions.

Lastly, the marked point bootstrap can be performed relatively quickly
compared to block bootstrap or subsampling. The marks that are
associated with the points are 
part of the actual estimator and are already computed in the
estimation step. Resampling using the marked point bootstrap only 
involves identifying which points are resampled with each resampling
region, and keeping track of how many times each point is
resampled. Inter-point distances and edge correction weights do not
have to be recalculated. With $N$ data points and $B$ bootstrap
samples, block bootstrap will take roughly $BN^2$ computations for a
statistic involving pairs of points. The marked point bootstrap will
involve roughly $N^2 + BN$ computations. The difference will be more
marked for three-point computations.

Simulation studies done in \citet{loh02a} showed that the empirical
coverages of confidence intervals obtained using the marked point
bootstrap can be much closer to the nominal 95\% level than those
obtained with block bootstrap or subsampling. 

We now describe how the marked point bootstrap can be used with
estimators of the two-point correlation function. The common
estimators of the two-point correlation function $\xi(r)$ are
\begin{eqnarray}
\hat{\xi}_{Nat}(r) & = & \frac{dd(r)}{rr(r)}-1, \label{eqn:nat} \\
\hat{\xi}_{DP}(r) & = & \frac{dd(r)}{dr(r)}-1, \label{eqn:dp} \\
\hat{\xi}_{Ham}(r) & = & \frac{dd(r)\cdot rr(r)}{dr(r)^2} -1, \label{eqn:ham}\\
\hat{\xi}_{Landy}(r) & = & \frac{dd(r)-2dr(r)}{rr(r)}+1, \label{eqn:landy} \\
\hat{\xi}_{Hewett}(r) & = & \frac{dd(r)-dr(r)}{rr(r)}, \label{eqn:hewett}
\end{eqnarray}
which are, respectively, the natural estimator \citep{kerscher2000}, and
estimators due to \citet{davis83, hamilton93, landy93, hewett82},
where $r$ is  
some distance of interest. In these expressions, $dd(r) = DD(r)/N^2, dr(r) =
DR(r)/NN_R$ and $rr(r)=RR(r)/N_R^2$, where $DD(r)=\sum_{x\in D} \sum_{y\in D: y\ne
  x} 1\{ |x-y| \in (r-dr, r+dr)\}/N^2$, 
$DR(r)=\sum_{x\in D} \sum_{y\in R} 1\{ |x-y| \in (r-dr, r+dr)\}/NN_R$
and $RR(r)=\sum_{x\in R} \sum_{y\in R: y\ne x} 1\{ |x-y| \in (r-dr, r+dr)\}/N_R^2$,
$R$ is a set of randomly generated points
(i.e.\ Poisson) in the observation region $A$, and $N$ and $N_R$ are
respectively the number of points in the real and random data sets.

To apply the marked point bootstrap, assign to
each point $x$ of the dataset marks $m_{x,1}=\sum_{y\in D: y\ne
  x} 1\{ |x-y| \in (r-dr, r+dr)\}$ and $m_{x,2}=\sum_{y\in R} 1\{
|x-y| \in (r-dr, r+dr)\}$. Bootstrap proceeds by resampling blocks of
points and recording the marks associated with them. For a bootstrap
sample, $x^*_j, j=1, \ldots N^*$, we then have
$$DD^*(r) = \sum_{j=1}^{N^*} m_{x^*_j,1}, \qquad
DR^*(r) = \sum_{j=1}^{N^*} m_{x^*_j,2},$$
and bootstrap estimates of the two-point correlation functions are
then obtained by substituting the above into
(\ref{eqn:nat})-(\ref{eqn:hewett}).
If each point $x_i$ of the actual data is resampled $n_i^*$ times, so
that $N^* = \sum_i n_i^*$, $DD^*(r)$ and $DR^*(r)$ can also be written
as
$$DD^*(r) = \sum_{i=1}^N (n_i^* \times m_{x_i,1}), \qquad DR^*(r) =
\sum_{i=1}^N (n_i^* \times m_{x_i, 2}).$$

Note that $RR$ does not need to be resampled, since it is used as an
approximation to an integral and has nothing to do with the actual
data. If, as is usually the case, estimation of $\xi(r)$ is needed for a
range of values of $r$, then the marks $m_{x,1}$ and $m_{x,2}$ would
be vectors, containing the relevant values for each value of $r$.

Estimators of the three-point correlation function can be bootstrapped
in a similar way. For example, an estimator of the three-point
correlation function is
\begin{eqnarray}
\zeta & = & \frac{ddd-ddr}{rrr} + 2, \label{eqn:3ptest}
\end{eqnarray}
introduced by \citet{peebles75}, where $ddd = DDD/N^3, ddr =
DDR/N^2N_R$ and $rrr = RRR/N_R^3$ and $DDD, DDR, RRR$ are counts
of triplets of points with the desired configuration, $DDD$ with all
points from the real data set and so on. The contribution to $DDD$
by any particular triplet of points is divided by 3 and assigned as
marks to each of the three points. For any individual point, all
these marks are summed together. For $DDR$, the contribution by each
triplet is
divided by 2 and assigned to the two real
data points. Bootstrap proceeds by resampling the real data points and
the values of $DDD^*$ and $DDR^*$ found by 
adding the marks of the resampled points.
Substituting these into
(\ref{eqn:3ptest}) gives the bootstrap estimate. Other similar
estimators, such as the three-point 
estimator of \citet{jing98} or the $N$-point estimators of
\citet{szapudi98}, can be bootstrapped in the same way.

\section{Simulation study}
\label{sect:simstudy}

We performed a simple simulation study to compare the performance of
confidence intervals obtained using the marked point bootstrap with
those obtained using normal approximations with Poisson errors,
varying the observation region size, number density and point process
model. For computational simplicity, we restrict to two dimensions.
We also performed an additional study with a large observation region
and approximately 50,000 points, showing the applicability of the
marked point bootstrap to datasets of size comparable to current
astronomy datasets. 

We used the Poisson point process model and a Neyman-Scott model to
generate the data points. The Neyman-Scott model is of historical
interest in astronomy as a model for galaxies \citep{neyman52}. It is
still commonly 
used to model point data in other fields \citep{diggle03, waag07}. We
chose the Neyman-Scott model as it is a model for clustered data with
closed-form expressions for the two-point correlation
function. The Neyman-Scott point datasets that we used are generated as
follows: parent points are distributed as a Poisson point
process with intensity $\lambda_p$. A Poisson number 
with mean $m$ of offspring points are then randomly scattered 
about each parent point. The collection of offspring points form the
point process. We set the dispersion function of offspring points
about parent points to be a bivariate normal density centered at the parent
point, with standard deviation $\sigma$. This specific Neyman-Scott
model is sometimes referred to as the modified Thomas model
\citep{stoyan95}. The 
two-point correlation 
function, $\xi(r)$, is zero for the 
Poisson model, while
$$\xi(r) = \frac{1}{4\pi \lambda_p \sigma^2}\exp\left\{
-\frac{r^2}{4\sigma^2} \right\}$$ for the modified Thomas model.
Thus the point pattern from a modified Thomas model is more clustered
if $\lambda_p$ or $\sigma$ is smaller. The quantity $\sigma$ also
controls the range of the correlation, with the range larger for
larger values of $\sigma$. We used several
values for $\lambda_p, m$ and $\sigma$ in our simulation study.

\begin{figure}
\begin{center}
\plotone{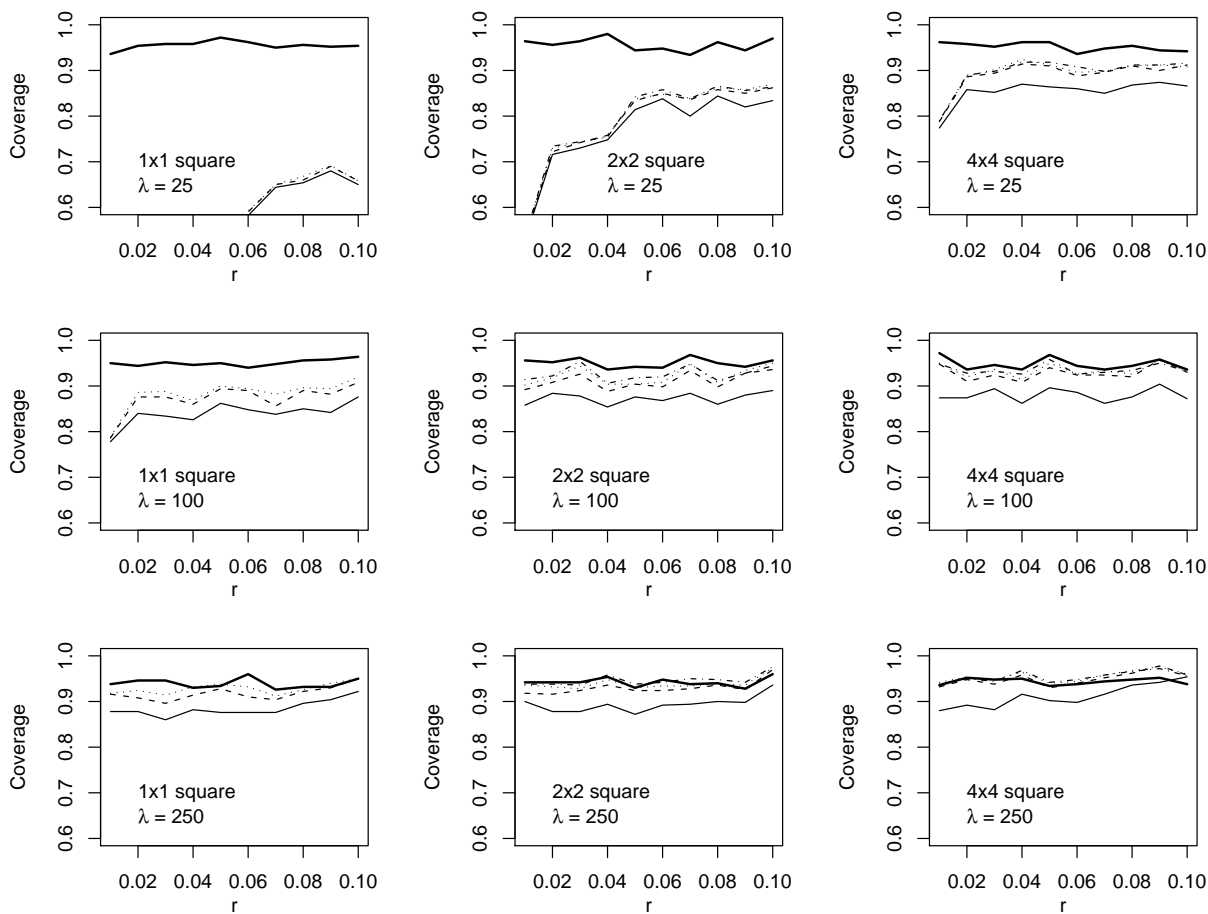}
\caption{Plots of the empirical coverage of nominal 95\% confidence
  intervals of the two-point correlation function for the Poisson
  point process model. The estimator used is that of
  \citet{hamilton93}. Confidence intervals are obtained using normal
  approximation with Poisson errors (thick solid line) and with the
  marked point bootstrap using different resampling block sizes.
 The block sizes were squares of
lengths 0.5, 0.33 and 0.25 for the 1 by 1 regions, of lengths 1,
0.67, 0.5, 0.33 for the 2 by 2 regions and of lengths 2, 1, 0.67,
0.5 for the 4 by 4 regions (solid, dashed, dotted and dashed-dotted
lines respectively for increasingly smaller blocks).
}
\label{fig:HamPoiCoverage}
\end{center}
\end{figure}

\begin{figure}
\begin{center}
\plotone{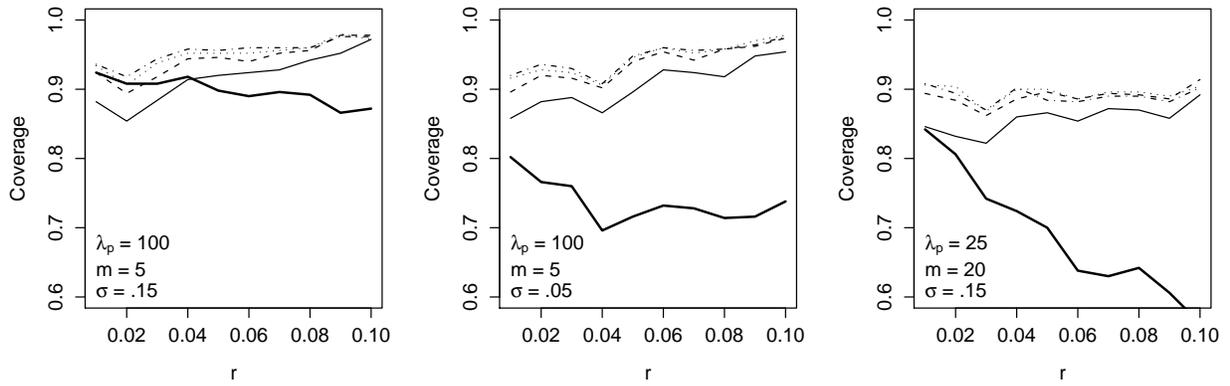}
\caption{Plots of the empirical coverage of nominal 95\% confidence
  intervals of the two-point correlation function for the modified 
  Thomas process model for realizations in a $2\times 2$ square. The
  estimator used is that of 
  \citet{hamilton93}. Confidence intervals are obtained using normal
  approximation with Poisson errors (thick solid line) and with the
  marked point bootstrap using different resampling block sizes (see text).}
\label{fig:HamTomCoverage}
\end{center}
\end{figure}

\begin{figure}
\begin{center}
\plotone{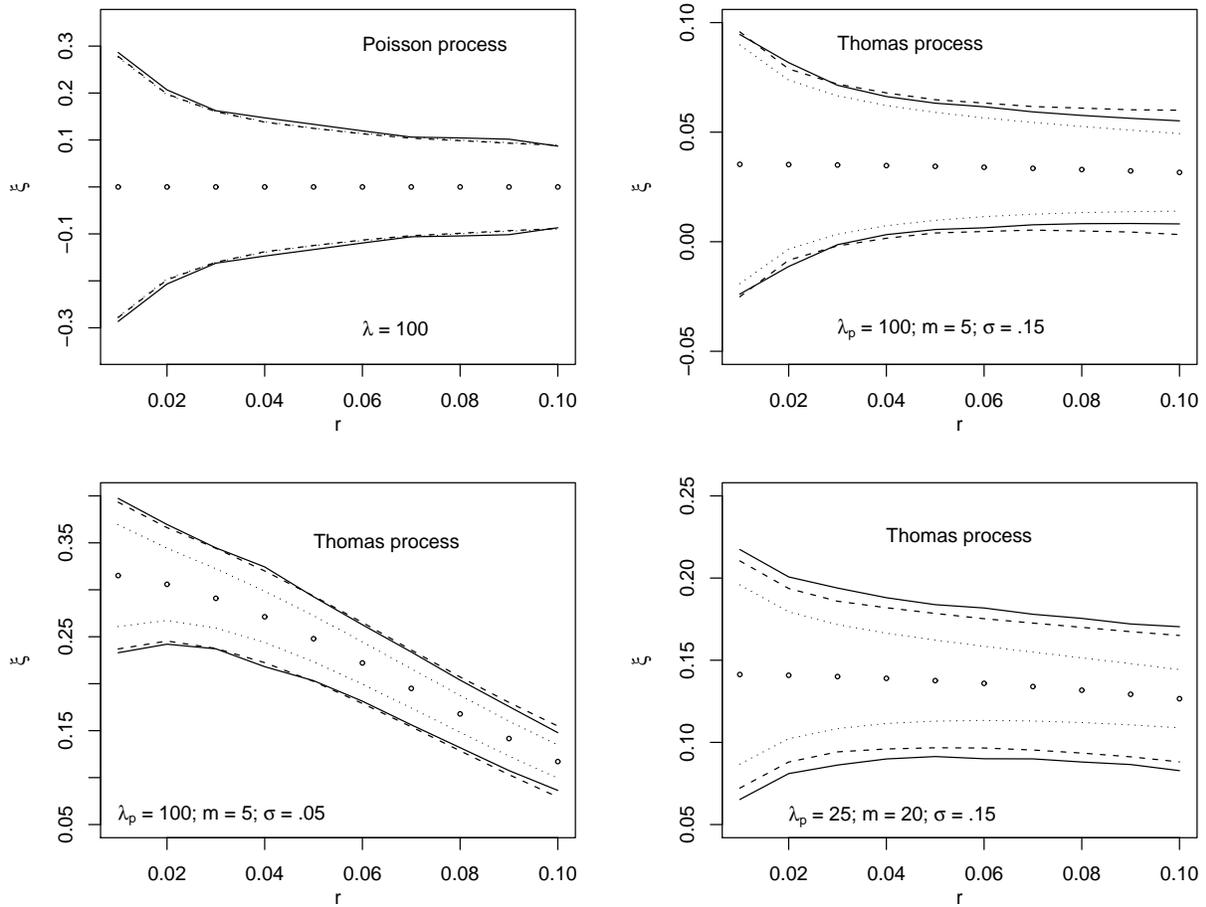}
\caption{Plots showing the true (solid), Poisson (dotted) and
  bootstrap (dashed) errors in estimates of $\xi$ for 500 sets
  of data simulated in a 2 by 2 square region using each of various
  point models. The true errors are obtained from the variability in
  the estimates of $\xi$ over the 500 data sets. For each data set,
  Poisson and bootstrap errors are computed. The errors shown in the
  plots are the average over the 500 data sets.}
\label{fig:Booterrors}
\end{center}
\end{figure}

For each point process model, we generated 500 realizations on the
unit square. For each
realization, we estimated $\xi(r)$ for $r=0.01, \ldots ,
0.1$. Bootstrap estimates were then produced from each realization
and a nominal 95\% confidence interval constructed. Thus for each
point process model, we have 500 95\% confidence intervals. We then
checked the the empirical coverage, i.e.\ the proportion of these that
contained the true value of 
$\xi(r)$, with proportion closer to 95\% being desirable. We also
constructed 500 confidence intervals using the normal approximation with
Poisson errors. The Poisson error $e_p$ is the inverse of the pair
counts for an uncorrelated data set of the same size as the actual data, as
given by \citet{landy93}.  The
95\% confidence intervals for $\xi$ based on the normal approximation
are thus given by $(\hat{\xi}\pm 
2e_p$). We then found the empirical coverage of these confidence
intervals. We then repeated the procedure for the $2\times 2$ and
$4\times 4$ squares. The results are summarized in Figures
\ref{fig:HamPoiCoverage} to \ref{fig:Booterrors}.

Figure \ref{fig:HamPoiCoverage} shows plots of the empirical coverage
of nominal 95\% confidence intervals of the two-point correlation
function for the Poisson process model, using the \citet{hamilton93}
estimator. Simulation results for the other estimators are similar and are not
shown. The thick solid lines in the plots show the empirical coverage
of confidence intervals obtained using normal approximation with
Poisson errors. Note that Poisson errors are correct in this case and
we find that the empirical coverage is close to 95\% for all the point
densities and observation region sizes considered.

The thin lines represent the empirical coverage of confidence
intervals obtained from the marked point bootstrap, with the different
line types representing different resampling block sizes. These were
squares of
lengths 0.5, 0.33 and 0.25 for the 1 by 1 regions, of lengths 1,
0.67, 0.5, 0.33 for the 2 by 2 regions and of lengths 2, 1, 0.67,
0.5 for the 4 by 4 regions (solid, dashed, dotted and dashed-dotted
lines respectively for increasingly smaller blocks). The difference
due to the block size used for resampling appear to be small. As
mentioned, this was an advantage of the marked point
bootstrap. \citet{loh02a} found greater variation of performance with
block size for subsampling and block bootstrap.

Compared with the Poisson empirical coverage, we find that at low
densities and smaller observation region sizes (plots towards the
upper left of Figure \ref{fig:HamPoiCoverage}), the bootstrap method
does poorly. However, the empirical coverage of the bootstrap
confidence intervals quickly increases towards 95\% with increasing
density (down the columns in Figure \ref{fig:HamPoiCoverage}) and/or
observation region size (across the rows in Figure
\ref{fig:HamPoiCoverage}), i.e. with larger sample sizes.  

\clearpage
\clearpage

\clearpage
\clearpage

\clearpage
\clearpage

\clearpage
\begin{figure}
\begin{center}
\caption{Plots showing sample realizations of the modified Thomas
  model corresponding to four different sets of parameter values,
  simulated on a $20\times 20$ square. The degree of clustering is
  higher in the top row, while the range of
clustering is larger in the right column.}
\label{fig:r20Thomas}
\end{center}
\end{figure}

\begin{figure}
\begin{center}
\plotone{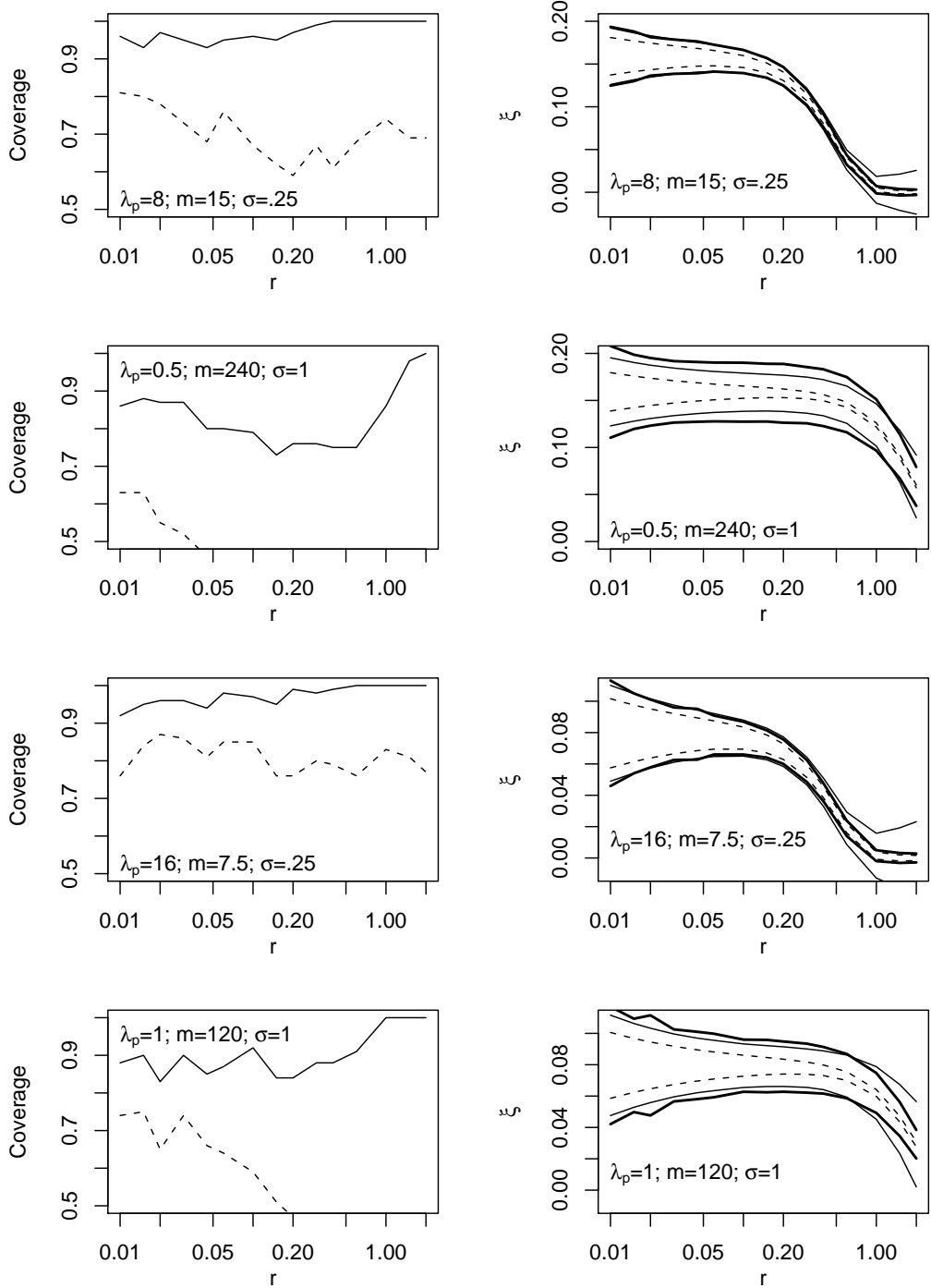}
\caption{Plots of the coverage (left) and errors (right) for the
  bootstrap (solid) and Poisson (dashed)
  methods, based on 100 
  simulated realizations from the Thomas model on the
  $20\times 20$ square. The thick solid lines in the plots on the
  right column represent the true errors.}
\label{fig:r20simulation}
\end{center}
\end{figure}
\clearpage

The top left plot of Figure \ref{fig:Booterrors} shows the Poisson
errors and bootstrap errors for the Poisson point process model
simulated on the $2\times 2$ square. The bootstrap errors shown in
this figure are from resampling with $0.33 \times 0.33$ squares. Also
shown in the plot are the true errors
as obtained from the estimates from 500 realizations. Notice that both
the Poisson and bootstrap errors are close to the true errors.

Figure \ref{fig:HamTomCoverage} shows similar plots for various
modified Thomas models, each with number density 500. 
The general behavior with increasing
observation region size for the Poisson model occurs here as
well. Thus to reduce the number of plots, we only include plots for the 2
by 2 observation regions, and show the relative performance of the
Poisson and bootstrap confidence intervals.

We find that when the point pattern is only weakly clustered (left
plot, for the case $\lambda_p=100, m=2.5$ and $\sigma=0.15$),
the Poisson confidence intervals had empirical coverage close to the
nominal 95\% level. However, as the other two plots in Figure
\ref{fig:HamTomCoverage} show, the empirical coverages of the Poisson
confidence intervals become lower than 95\% as the degree and/or range
of clustering increases (i.e.\ with smaller $\sigma$ or $\lambda_p$). 
On the other hand, the boostrap confidence intervals attain coverage
much closer to 95\% for all the cases shown, regardless of the degree
of clustering. Plots of the Poisson and bootstrap error estimates are
shown in Figure \ref{fig:Booterrors}. Notice that the Poisson
approximation underestimates the true error as the degree of
clustering increases, while the bootstrap error estimates remain close
to the true errors, even for the modified Thomas model.

Thus we find that the performance of the Poisson confidence intervals
is sensitive to the degree of clustering of the point pattern. If the point
pattern is Poisson, or weakly clustered, the empirical coverage of
Poisson confidence intervals is
close to the nominal level, even with small sample sizes. However,
performance quickly deteriorates with greater degree of clustering.
On the other hand, the bootstrap confidence interval does not perform
well with small sample sizes. With moderate sample
sizes, however, the bootstrap method performs rather well, over a wide
range in the degree of clustering. 

We performed an additional set of simulations using data sets of
roughly 50,000 points in a  $20\times 20$
square and estimating $\xi(r)$ for $r= 0.01$ to 2. Other than the
restriction to 2D, the data size 
and range of $r$ is roughly of the scale found in current astronomy
data. We used the modified Thomas model and chose four sets of
parameter values, varying the degree and range of clustering but
with the same number density. A sample
realization from each of the four models is shown in Figure
\ref{fig:r20Thomas}. The models corresponding to the top row in Figure
\ref{fig:r20Thomas} have higher clustering than the models on the
bottom row. For models in the same row, the strength of clustering is
similar, but the model on the right has a longer correlation range.
We used square resampling blocks of side length 5, 2.5 and 2 and
results were very similar.

The results are summarized in Figure \ref{fig:r20simulation}, which
show the empirical coverage of confidence intervals (left column) and
errors (right column) obtained from the marked point bootstrap and
with Poisson errors, for each of the four Thomas models.
The plots qualitatively show the same relative
performance between Poisson errors and bootstrap as found in the
earlier simulation study. When the range of
clustering is large, the empirical coverage of confidence intervals
based on Poisson errors and the normal approximation is very low
(second and fourth plots on the left column of Figure
\ref{fig:r20simulation}). The
coverage of the bootstrap intervals are affected too, but by much less.
When the correlation is large, the Poisson errors substantially
under-estimate the true errors, while the marked bootstrap errors were
more realistic. At the larger values of $r$, especially when $\xi$ is
near 0, the bootstrap procedure appears to be somewhat conservative,
while the Poisson errors become more accurate.

We made some time
measurements of various sections of the algorithm: the functions
computing $DD$ and $DR$ took 1 minute and 13 minutes
respectively. Here, $N_R = 200,000$ and we did not use any
sophisticated methods (such as tree-based algorithms) to speed up the
computation. The bootstrap function, generating 999 samples and
computing the estimates, took roughly 1 minute, showing the
feasibility of the procedure for large data sets.  The speed of the
marked point bootstrap is due to the fact that the marks that are
resampled have already been computed as part of the estimation. The
additional computational burden of the bootstrap is sampling the
points and keeping track of the number of times each point is
resampled.

\section{Discussion}
\label{sect:discussion}

In this paper, we introduced the marked point bootstrap as a
method to bootstrap spatial data for estimating errors without
specific model assumptions. In particular,
we described how the method can be applied to estimators of the two-
and three-point correlation functions.
With the non-parametric bootstrap, errors are obtained from the actual
data. There is no need choose a model, select parameter values or
generate mock catalogs using $N$-body simulations. Thus errors
obtained from non-parametric bootstrap can be used to compare with
errors obtained from other methods with more specific
model assumptions.

For non-parametric spatial bootstrap, we propose the marked point bootstrap
over the more common block bootstrap or subsampling methods. There are
several advantages of the marked point bootstrap.
Firstly, by using information from actual pairs
or triplets of points in the data, bootstrap confidence intervals
using the marked point bootstrap attain better empirical coverage than
confidence intervals constructed using block bootstrap (see
\citealt{loh02a} for a comparison of these two methods). 

Secondly, in
the marked point bootstrap, it is the marks that are used to compute
the bootstrap estimate. Thus the resampled points do not have to be
arranged to form a new point pattern. This makes it a lot easier to
bootstrap data that are observed in irregularly shaped regions that are
common in astronomy. In \citet{loh02a} for example, bootstrap on an
absorber catalog was done using slices as well as spheres, with
similar results for both types of resampling regions.

Thirdly, the marks used for resampling are part of the original
estimate and are computed during the estimation
step. The only additional
computation required by the marked point bootstrap  involves selecting
points (that is, testing whether each point lies in a resampling region
or not), and keeping track of the number of times each point is
resampled. Unlike the block bootstrap, there is no need to 
re-compute from scratch the estimates for each bootstrap sample.
This difference in computation is even greater for higher-order
statistics. These properties of the marked point bootstrap make it a
computationally feasible tool for analysis.

Our study here suggests that
non-parametric bootstrap can yield valid estimates of errors under a
wide range of point patterns. The lack of specific model assumptions
means that the non-parametric bootstrap method, and in particular the
marked point bootstrap, can serve as an alternate and complementary method for 
quantifying errors. Having estimates of errors obtained using Poisson
approximations, parametric and non-parametric bootstrap allows one to
have a better sense of the size of errors involved in an analysis.

The simulation study performed here shows that bootstrap confidence
intervals do attain coverage close 
to the nominal level, even for the clustered point patterns
where Poisson errors are known to be inaccurate, when sample sizes are
large. More specifically, bootstrap performance improves with
increasing number density, and also with increasing observation region
size relative to the correlation length. 
Unfortunately, in astronomy, the correlation length may be of the same
scale as the observation region. If the values of $r$ at which the
correlation function estimates are computed are small relative to the
resampling blocks (and the observation region), then although the
bootstrap procedure would distort the dependence structure at the
large scales, it would still be valid for these smaller values of $r$.

If, instead,
$\xi(r)$, say, for $r$ close to the size of the observation region
is of interest, then the bootstrap procedure would start to
break down, in the sense that the empirical coverage of confidence may
not be close to the nominal level, and the bootstrap errors not
reflect the true errors. In this case, the amount of information
contained in the data is smaller and the boundary effects are
magnified. With respect to the marked point bootstrap, larger blocks
are needed to capture the dependence structure at this larger
scale. For a fixed sample, 
this cannot be done without reducing the variability of the bootstrap samples.
The method that might work best is parametric bootstrap, assuming that
the model is correct, and that the parameter values used are close to
the true values. 
Non-parametric bootstrap can still be useful here. Firstly, it is at
least a better choice than Poisson errors, since the latter would
grossly underestimate the true errors. Secondly, it can provide additional
error estimates to compare with the errors obtained with the assumed
model. For these most challenging instances, having a variety of
methods can only be beneficial.

\acknowledgments

This research is supported in part by
National Science Foundation award AST-0507687.

\end{document}